\documentclass[english,11pt,aps,prd,a4paper,preprintnumbers,floatfix,nofootinbib,showpacs,superscriptadress, notitlepage]{revtex4-2}

\usepackage{graphicx}
\usepackage[utf8]{inputenc} 
\usepackage{amsmath}
\usepackage{amssymb}
\usepackage{float}
\usepackage{comment}
\usepackage{slashed}
\usepackage[normalem]{ulem}
\usepackage{cancel}
\usepackage{dirtytalk}
\usepackage{empheq}

\newcommand*\widefbox[1]{\fbox{\hspace{2em}#1\hspace{2em}}}

\makeatletter 
\renewcommand{\fnum@figure}{\textbf{Fig.~\thefigure}}
\makeatother
\usepackage{multirow}
\usepackage{rotating}
\usepackage[usenames,dvipsnames]{color}
\usepackage[colorinlistoftodos]{todonotes}
\usepackage[colorlinks=true,citecolor=blue,urlcolor=blue, pdfborder={0 0 0}]{hyperref}
\usepackage[normalem]{ulem}
\definecolor{darkred}{rgb}{0.6,0,0}
\usepackage[colorinlistoftodos]{todonotes}
\definecolor{linkcolor}{rgb}{0,0,0.5}

\usepackage{soul}
\definecolor{mightnightblue}{RGB}{25,25,112}

\definecolor{brown}{rgb}{0.59, 0.29, 0.0}

\begin{document}
\title{\color{BrickRed} Neutrino millicharge and other electromagnetic interactions with COHERENT-2021 data}
\author{Amir N.\ Khan}
\email{amir.khan@mpi-hd.mpg.de}
\affiliation{Max-Planck-Institut f\"ur Kernphysik, Postfach 103980, D-69029
Heidelberg, Germany}

\begin{abstract}%
We analyze new data from the COHERENT experiment of the coherent neutrino-nucleus scattering to investigate the electromagnetic interactions of neutrinos. With almost double the statistics and precision now, the statistical significance of the observed process has now enhanced to 11.6$\sigma$. We derive constraints on the electromagnetic properties of neutrinos using the new COHERENT data. The constraints improve by more than a factor of two compared to the previous bounds. Furthermore, we discuss the unique behavior of the neutrino millicharge at lower energy recoils and show its unique dependence on its interference with the standard model contribution, inverse power of recoil energy and the mass of the target particle in comparison to the other interactions.
\end{abstract}
\date{\today }
\pacs{xxxxx}
\maketitle
\section{Introduction}
Coherent elastic neutrino-nucleus scattering (CE$\nu$NS) is a SM process that was predicted forty years ago \cite{Freedman:1973yd, Freedman:1977xn,Tubbs:1975jx,Drukier:1984vhf} and was recently observed by the COHERENT experiment \cite{Akimov:2017ade, Akimov:2018vzs}. The importance of the process ranges from its ability as a precision probe of the SM parameters \cite{Freedman:1973yd, Freedman:1977xn, Tubbs:1975jx, Drukier:1984vhf, Barranco:2005yy, Lindner:2016wff,Cadeddu:2017etk, Arcadi:2019uif,AristizabalSierra:2019zmy,Papoulias:2019lfi} to test new interactions at low momentum transfer including its importance for the direction detection of dark matter \cite{Anderson:2012pn,deNiverville:2015mwa,Ge:2017mcq,Coloma:2017ncl,Bauer:2018onh, Billard:2018jnl,Denton:2018xmq,Dutta:2019eml,Kosmas:2017tsq,Khan:2021wzy,Tomalak:2020zfh,Coloma:2022avw}. The COHERENT collaboration has updated the result for the aforementioned process by doubling the statistics and the precision by reducing the overall systematic errors to half \cite{COHERENT:2021xmm}. In particular, the error on the quenching factor improves from 25\% to 4\%. Compared to the first result at 6.7$\sigma$, the updated significance level has now reached 11.6$\sigma$. With this improvement, it is natural to expect a better sensitivity to any new physics or improvement in the limits. To this aim, we analyze the new data to constrain the neutrino electromagnetic properties if they contribute to the CE$\nu$NS and derive constraints on neutrino magnetic moment, millicharge, charge radius and neutrino anapole moment using the new COHERENT data.

In low energy scattering experiments with low target recoils, the sensitivity to any new physics mainly depends on three factors: $(i)$ whether the new physics interactions interfere with the SM interactions or not, $(ii)$ the proportionality of the new physics coupling strength to the inverse power of the target's recoil energy, and $(iii)$ mass of the target particle. These three factors are different for the four types of electromagnetic interactions in the $\nu-$e elastic scattering and in the CE$\nu$NS processes, as we will discuss later. For example, the neutrino millicharge is more sensitive to low energy recoils \cite{Khan:2020vaf, Khan:2020csx, XENON:2020rca} because it interferes with the SM interactions, and its coupling is proportional to the inverse square of the target recoil energy. We will discuss this aspect in more detail in section V. These kinematical considerations are equally valid for the target recoils in the dark matter scattering \cite{Knapen:2020aky}. 

We organize the rest of the paper as follows. In the next section, we discuss the basics of the differential cross-section of the CE$\nu$NS in the SM. Then, in Sec. \ref{sec:analysis}, we discuss data analysis for the COHERENT setup with the new data. Then, in Sec. \ref{sec:EMprorties}, we introduce the electromagnetic properties of neutrinos and derive constraints using the new COHERENT data. Next, in Sec. \ref{sec:Kineconside}, we discuss in detail why millicharge neutrinos are kinematically more special than the other electromagnetic properties of neutrinos. Finally, we provide the conclusion of this work in sec \ref{sec:concl}. 
\section{Coherent Elastic neutrino nucleus scattering}\label{sec:formalism}
At the tree level in the SM, the differential cross-section 
of the neutrino with flavor ‘$\alpha$’ scattering off the spin-0 nucleus of CsI with proton number ‘$Z$’ and neutron number ‘$N$’ is given by ~\cite{Freedman:1973yd, Freedman:1977xn, Tubbs:1975jx, Drukier:1984vhf, Barranco:2005yy, Lindner:2016wff}, 
\begin{equation}
\frac{d\sigma_{\alpha}}{dT}(E_{\nu},T) = \frac{G_{F}^{2}M}{\pi }
\left[Zg_{p}^{V}+ Ng_{n}^{V})\right]^{2}
\left( 1-\frac{T}{E_{\nu}}-\frac{MT}{2E_{\nu }^{2}}\right) F^{2}(q^{2})\,,
\label{eq:diff-crossec}
\end{equation}%
where ‘$G_{F}$’ is the Fermi constant, ‘$E_\nu$’ is the energy of the incoming neutrinos, 
‘$T$’ is nuclear the recoil energy,  $q^2=2 M T$ is the squared momentum transfer, and ‘$M$’ is the mass of the target nucleus. Here, $g_{p}^{V}=(2g_{u}^{V}+ g_{d}^{V})$ and $g_{n}^{V}=(g_{u}^{V}+2g_{d}^{V})$, where $g_{u}^{V}$ and $g_{d}^{V}$ are the neutral current coupling constants for the ‘up’ and ‘down’ quarks which, in terms of the weak mixing angle ‘$\theta _{W}$’ at tree level are given by
\begin{eqnarray}
g_{u}^{V}& = & \frac{1}{2}-\frac{4}{3} \sin ^{2}\theta _{W}\,,\\
g_{d}^{V}& = & -\frac{1}{2} +\frac{2}{3}\sin ^{2}\theta _{W}\,.
\label{eq:gv&ga}
\end{eqnarray}%
We will use $\rm sin ^{2}\theta_{W}=0.23857 \pm 0.00005$, the low energy $(q \rightarrow 0)$ value evaluated in $\overline{\rm MS}$ scheme \cite{ParticleDataGroup:2020ssz}. Using the low energy value of $\rm sin ^{2}\theta_{W}$ and including the small radiative corrections \cite{Erler:2013xha,ParticleDataGroup:2020ssz}, we find the values of the coupling constants $g_{u}^{V} = 0.197, \ \ g_{d}^{V} = -0.353$ for electron neutrinos and $g_{u}^{V} = 0.191, \ \ g_{d}^{V} = -0.350$ for muon neutrinos.
In Eq. (\ref{eq:diff-crossec}), $F(q^{2})$ is the nuclear form factor, and we use the Klein-Nystrand form \cite{Klein:1999gv} as given in the following
\begin{equation}
F(q^{2})=\frac{4\pi \rho _{0}}{Aq^{3}}[\sin (qR_{A})-qR_{A}\cos (qR_{A})]%
\left[ \frac{1}{1+a^{2}q^{2}}\right],  \label{F-bessel}
\end{equation}%
where $\rho _{0}$ is the normalized nuclear number density, $A$ is the  atomic number of CsI, $%
R_{A}=1.2A^{1/3}\, \mathrm{fm}$ is the nuclear radius, and $a=0.7\, \mathrm{fm
 }$ is the range of the Yukawa potential. 
\section{Data Analysis}\label{sec:analysis}
The COHERENT detector receives a prompt signal from the mono-energetic (29.8 MeV) beam of muon-neutrinos $(\nu _{\mu })$ produced from the $\pi^+$ decay at rest ($\pi ^{+}\rightarrow \mu ^{+}\nu _{\mu }$) at the Oak Ridge Spallation Neutron Source. Subsequently, continuous fluxes of electron-neutrinos ($\nu _{e}$) and muon-anti-neutrinos ($\bar{\nu}_{\mu })$ with energy peaks, respectively, around 35 Mev and 52.8 MeV produced in $\mu^+$ decays ($\mu ^{+}\rightarrow \nu
_{e}e^{+}\bar{\nu}_{\mu }$) with the characteristic time scale of the muon lifetime, namely 2.2 $\mu$ sec, is received. The fluxes are produced from $3.20\times 10^{23}$ protons on target from the liquid mercury. The average production rate of the SNS neutrinos from the pion decay chain is $r=0.0848$ neutrinos of each flavor per proton \cite{COHERENT:2021xmm}.

The detector, located at a distance $L=19.3$ m from the source, uses CsI[Na] as a target, where the Na contributes small enough to be neglected. For such a setup, the total number of events of the nuclear recoil in a given energy bin ‘$i$’ and neutrino flavor ‘$\alpha$’ reads
\begin{equation} 
N_{\alpha}^i=N 
\int_{T^{\prime \rm i}}^{T^{\prime \rm i+1}}\hspace{-1.5em}dT^{\prime}
\int_{0}^{T^{\rm max}}\hspace{-1.5em}dT 
\int_{E_{\nu }^{\rm min }}^{E_{\nu }^{\rm max}}\hspace{-1.3em}dE_{\nu }
\frac{d\sigma_{\alpha } }{dT}(E_{\nu },T)\frac{%
d\phi _{\nu _{\alpha }}(E_{\nu })}{dE_{\nu }}\mathcal{E} (T^\prime) G( T^\prime, T),
\label{eq:eventrt}
\end{equation}%
where $\mathcal{E} (T^\prime)$ is the detection efficiency function, $G(T^\prime, T)$ is the gamma distribution function for the detector energy resolution, $T$ and $T^\prime$ denotes the nuclear recoil energy and the reconstructed recoil energy, respectively.
Here, $N=\left(2m_{\mathrm{det}}/M_{\rm CsI}\right) N_{A}$ is the total number of CsI nucleons, $m_{\mathrm{det}}=14.57$ kg, $N_{A}$ is the Avogadro's  number,  $M_{\rm CsI}$ is the molar mass of CsI, $E_{\nu }^{\min}= \sqrt{MT/2}$, $M$ is the mass of the target nucleus, $E_{\nu }^{\max }$ is the maximum neutrino energy and $d\phi _{\nu _{\alpha }}(E_{\nu })/dE_\nu$ is the flux corresponding to the flavor ‘$\alpha$’ \cite{Khan:2021wzy}.

The recent measurement of COHERENT \cite{COHERENT:2021xmm} considers the recoiled energy-dependent quenching factor, $f_{\rm q}(T^{\prime})$ and measures the energy spectrum in terms of photo-electrons (p.e). Therefore, to calculate the total number of events in a particular bin ‘i’ of photo-electrons, we use the following relation between the recoil energy and the number of photo-electrons ($N_{\rm p.e}$)
\begin{equation}\label{qf}
N_{\rm p.e.} = f_{q}(T^{\prime})\times T^{\prime}\times Y,
\end{equation}%
where $ \rm Y = 13.35 ~ \rm photons$/keV is the light yield and $f_{q}(T^{\prime})$ is taken from \cite{COHERENT:2021xmm}.

To fit the data of the energy spectrum in Fig. 3 of ref. \cite{COHERENT:2021xmm} with the model including the SM and the new physics parameters, we use the following least-square function 
\begin{equation}
\chi ^{2}=\underset{i=2}{\overset{9}{\sum }}
\left(\frac{N_{\rm obs}^{i}-N_{\rm exp}^{i}(1+\alpha)-B^{i}(1+\beta)}
{\sigma^{i}}\right)^2
+\left( \frac{\alpha }{\sigma _{\alpha }}\right) ^{2}+\left(\frac{\beta }{
\sigma_{\beta }}\right)^{2}\,,
\label{eq:chisq}
\end{equation}%
where $N_{\rm obs}^{i}$ denotes the observed events in the $i$-th energy bin and $\sigma ^{i}$ is the relevant uncertainty. $N_{\rm exp}^{i}$ is the total number of expected events, which is the sum of the three neutrino flavors as given in eq.\ (\ref{eq:eventrt}). $B^{i}$ is the sum of prior predicted beam-related neutron and the neutrino-induced neutron backgrounds in a given energy bin. The first and second penalty terms correspond to the systematic uncertainty of the signal and backgrounds where ‘$\alpha$’ and ‘$\beta$’ are the corresponding nuisance parameters. The uncertainty in the signal is $\sigma_{\alpha}$ = 0.127 and the uncertainty on the total background is $\sigma_{\beta}$ = 0.6. The signal uncertainty includes a contribution from the neutrino flux, quenching factor, efficiency, form factor and light yield. We took all information from ref. \cite{COHERENT:2021xmm}. 
One can expect stronger or even weaker constraints with timing information since the only parameter affected by the timing information is the total efficiency. However, since the time and recoil energy is completely uncorrelated and thus efficiencies for the two cases are also uncorrelated \cite{COHERENT:2021xmm}, our results are thus valid without incorporating the timing information. In order to avoid the possible flavor dependence on the timing information, we restrict our analysis to only flavor-conserving interactions.

\section{Electromagnetic interactions of neutrinos in CE$\nu$NS }\label{sec:EMprorties}
\subsection{Millicharge neutrinos}
The electromagnetic contribution due to the electrically charged neutrinos, parameterized in terms of $Q_{\alpha \alpha}$, to the SM weak interaction for the coherent neutrino-nucleus ($\nu-N$) scattering is given by the interactions
\begin{equation}
\mathcal{L}_{\alpha}^{em}=-ie
\left(Q_{\alpha \alpha}{\overline{\nu}_{\alpha}}\gamma_{\mu }{\nu}_{\alpha} + {\overline{N}}\gamma_{\mu}{N} \right) A^{\mu},
\label{eq:CHargN}
\end{equation}
where $A^{\mu}$ is mediating electromagnetic field and ‘$e$’ is the unit electric charge. Since the electromagnetic interaction terms add coherently to the vector part of the weak interaction, this modifies the weak mixing angle $\theta _{W}$ in eq. (\ref{eq:gv&ga}) accordingly as
\begin{eqnarray}
\sin ^{2} \theta _{W}\rightarrow \sin ^{2}\theta _{W} \left(1 - \frac{\pi \alpha_{em}}{\sqrt{2} \sin ^{2}\theta _{W} G_F MT}Q_{\alpha\alpha}\right)
\label{eq:milicharge}
\end{eqnarray}
where $\alpha_{em}$ is the fine structure constant.

We estimate the statistical significance of the millicharge neutrinos by fitting the two parameters, $Q_{ee}$ and $Q_{\mu \mu}$. We consider two cases while fitting parameters. First, we fit one parameter at a time and fix the other to zero, and we show the result in the left-side plot of Fig. \ref{fig:milchage}. Next, we fit the two parameters together and show the results in the right-side plot of Fig. \ref{fig:milchage}. One can see from Fig. \ref{fig:milchage} that because of the interplay between the SM and the neutrino millicharge contribution (interference effect) and the dependence on the inverse square of the nuclear recoil energy in the cross-section, the two-parameter fit prefers non-zero values at the best-fit minimum.
We obtain the following constraints from the one parameter fits at 90\% C.L.,

\begin{empheq}[box=\widefbox]{align}
-0.55\times 10^{-7}<Q_{\mu\mu}/e<0.75\times 10^{-7}\,, \nonumber \\
-1.10\times 10^{-7}<Q_{ee}/e<3.90\times 10^{-7}\,. 
\end{empheq}

\begin{figure}[t]
\begin{center}
\includegraphics[width=6.5in]{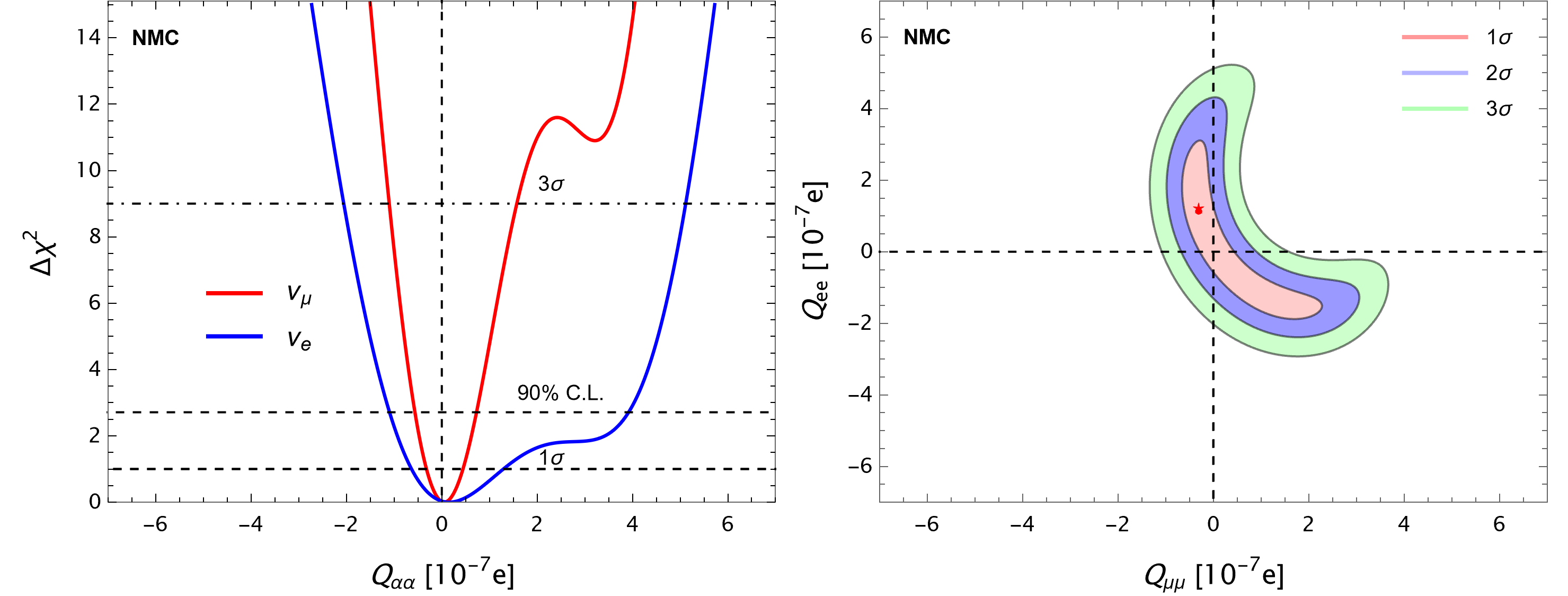}
\label{fig:milchage}
\end{center}
\caption{Neutrino millicharge (NMC) of the muon and electron flavors from the new COHERENT data for one (left) and two (right) parameter fits. The red star in the right-hand side plot corresponds to the best-fit value. See text for discussion.}
\label{fig:milchage}
\end{figure}

Stronger limits on millicharge neutrinos come from the observational studies \cite{Barbiellini:1987zz, Raffelt:1999gv, Davidson:2000hf,Melchiorri:2007sq,Studenikin:2012vi}. The strongest upper limit on the millicharge neutrino is $Q_\nu \leq 2 \times 10^{-15} e $ from the time arrival dispersion and the energy spread of neutrinos from SN1987A \cite{Barbiellini:1987zz}. The laboratory bounds from the $\nu-e$ are also several orders of magnitude smaller in size than the bounds of this study \cite{Davidson:1991si, Babu:1993yh,Bressi:2011yfa, Gninenko:2006fi,Chen:2014dsa,TEXONO:2018nir,Khan:2019cvi,Cadeddu:2019eta,Cadeddu:2020lky,Khan:2020vaf}. For instance, the TEXONO experiment derives the limit, $Q_\nu \leq 2.1 \times 10^{-12} e $. However, we can easily understand this difference from the kinematical considerations, as we will discuss in Sec. \ref{sec:Kineconside}. However, the robustness of the bounds depends on the experimental details. Exploring how the observational constraints also depend on the kinematic details of the astrophysical environments would be interesting.  

\subsection{Neutrino magnetic moment}
To understand the importance of interference and the low recoil energy dependence of the millicharge neutrinos, we also compare the neutrino magnetic moment to the same data. In general, for the Majorana ($M$) or Dirac ($D$) neutrinos' couplings to the electromagnetic field strength ($F^{\mu\nu}$), the magnetic moments appear as \citep{Fujikawa:1980yx, Shrock:1982sc, Vogel:1989iv,Abak:1989kp, Grimus:1997aa}
\begin{equation}
{\cal L}^M = -\frac 14 \bar \nu_{\alpha L}^c \, \lambda_{\alpha \beta}^M \, \sigma_{\mu\nu} \, \nu_{\beta L} \, F^{\mu\nu}~\mbox{ or }~{\cal L}^D =-\frac 12 \bar \nu_{\alpha R} \, \lambda_{\alpha \beta}^D \, \sigma_{\mu\nu} \, \nu_{\beta L}\, F^{\mu\nu}\,,
\label{eq:MM}
\end{equation}
where $\lambda^X = \mu^X - i \epsilon^X$, which is hermitian for the Dirac neutrinos and antisymmetric for Majorana neutrinos. Only transition magnetic moments are possible for Majorana neutrinos while the flavor diagonal is zero. Because of the unknown final state neutrino flavor in a scattering process, in practice, no distinction between Dirac and Majorana neutrinos is possible. For simplicity, we only consider the flavor diagonal cases for the electron ($\mu_{\nu e}$) and muon neutrinos ($\mu_{\nu \mu}$). In the SM, a non-zero neutrino magnetic moment can arise at the one-loop level, which quantifies as follows,  \citep{Fujikawa:1980yx} 
\begin{equation}
\mu_{\alpha \beta} = \frac{3eG_F m_{\nu}}{8 \sqrt{2} \pi^2} \sim 3 \times 10^{-19} \mu_{B}\left(\frac{m_\nu}{1 \rm eV}\right)
\end{equation}


As clear from eq. (\ref{eq:MM}), the helicity of the final state neutrino changes in interaction due to the magnetic moment. Therefore there is no interference with the SM cross-section, and the corresponding contribution adds to the standard model at the cross-section level. We add the following differential cross-section \cite{Khan:2019cvi} for the neutrino magnetic moment (MM) of neutrino scattering off a spin-0 nucleus to the SM cross-section in eq. (\ref{eq:diff-crossec}),
\begin{equation}
\frac{d\sigma_{\alpha}^{MM}}{dT}(E_{\nu},T) =\left( \frac{\pi \alpha_{em}^2\,\mu _{\nu \alpha}^2}{m_{e}^2}\right)%
\text{ }\left( \frac{1}{T}-\frac{1}{E_{\nu }}+\frac{T}{4E_{\nu }^{2}}\right) Z^2 F^{2}(q^{2}),    
\label{eq:EDM}
\end{equation}
where we take the magnetic moment ($\mu_{\nu \alpha}$) in units of Bohr's magneton ($\mu_B$), and $m_e$ is the electron mass. One can notice that compared to the millicharge of neutrinos as given in eq. (\ref{eq:milicharge}) in combination with eq. (\ref{eq:diff-crossec}), the neutrino magnetic moment has no interference with the SM, and the dependence on the inverse power of the nuclear recoil is only linear in the leading terms. 

\begin{figure}[t]
\begin{center}
\includegraphics[width=6.5in]{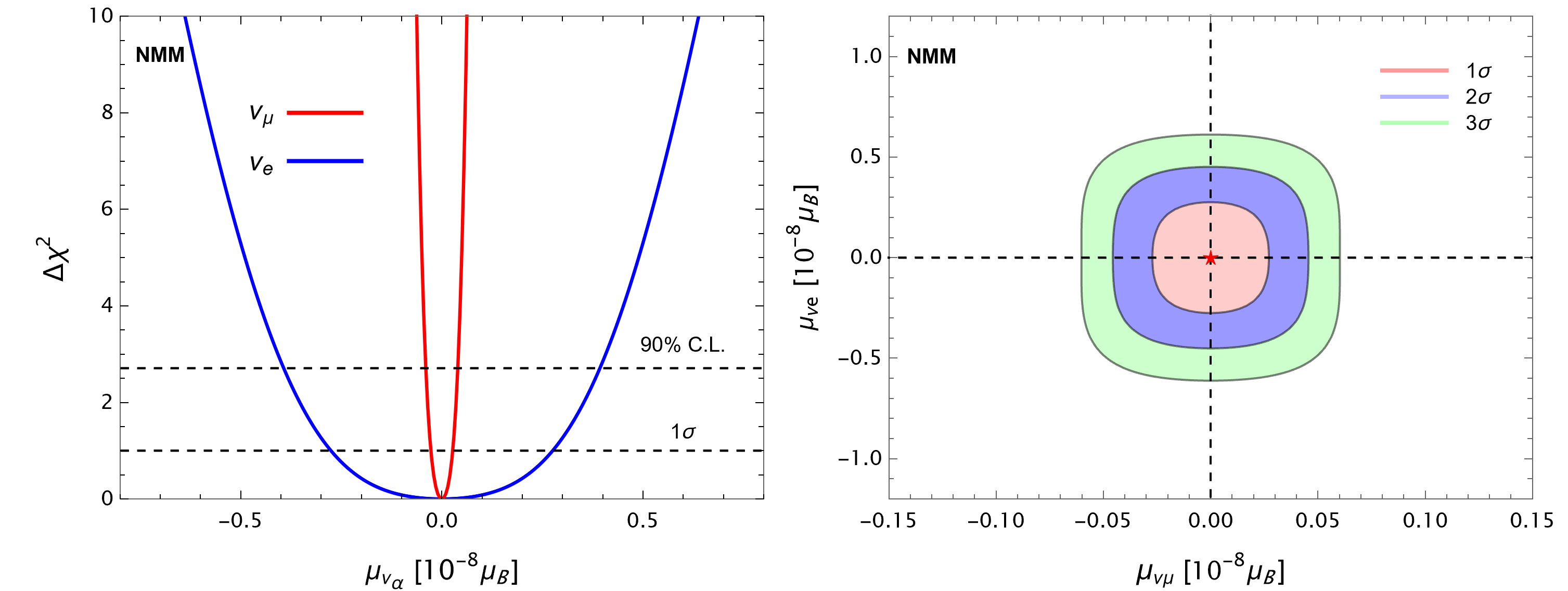}
\end{center}
\caption{Neutrino magnetic moment (NMM) of the electron and muon neutrino from the new COHERENT data for the one (left) and two (right) parameters fits. The red star in the right-hand side plot corresponds to the best-fit value. See text for discussion.}
\label{fig:DM}
\end{figure}

In this case, we consider two parameters $\mu_{\nu_{\mu}}$ and $\mu_{\nu_e}$ and fit them to the new COHERENT data \cite{COHERENT:2021xmm} using eq. (\ref{eq:chisq}), first, with one parameter at a time while keeping the other zero and then fitting two parameters together. The results for both cases are shown respectively in the left-side and right-side plots of Fig. \ref{fig:DM}. We obtain the following constraints from the one-parameter fits at 90\% C.L.,

\begin{empheq}[box=\widefbox]{align}
-0.04\times 10^{-8} <\mu_{\nu_\mu}/\mu_B<0.04\times 10^{-8}\,, \nonumber \\
-0.40\times 10^{-8} <\mu_{\nu_e}/\mu_B<0.40\times 10^{-8}\,.  
\end{empheq}

\subsection{Neutrino charge radius}

\begin{figure}[t]
\begin{center}
\includegraphics[width=6.5in]{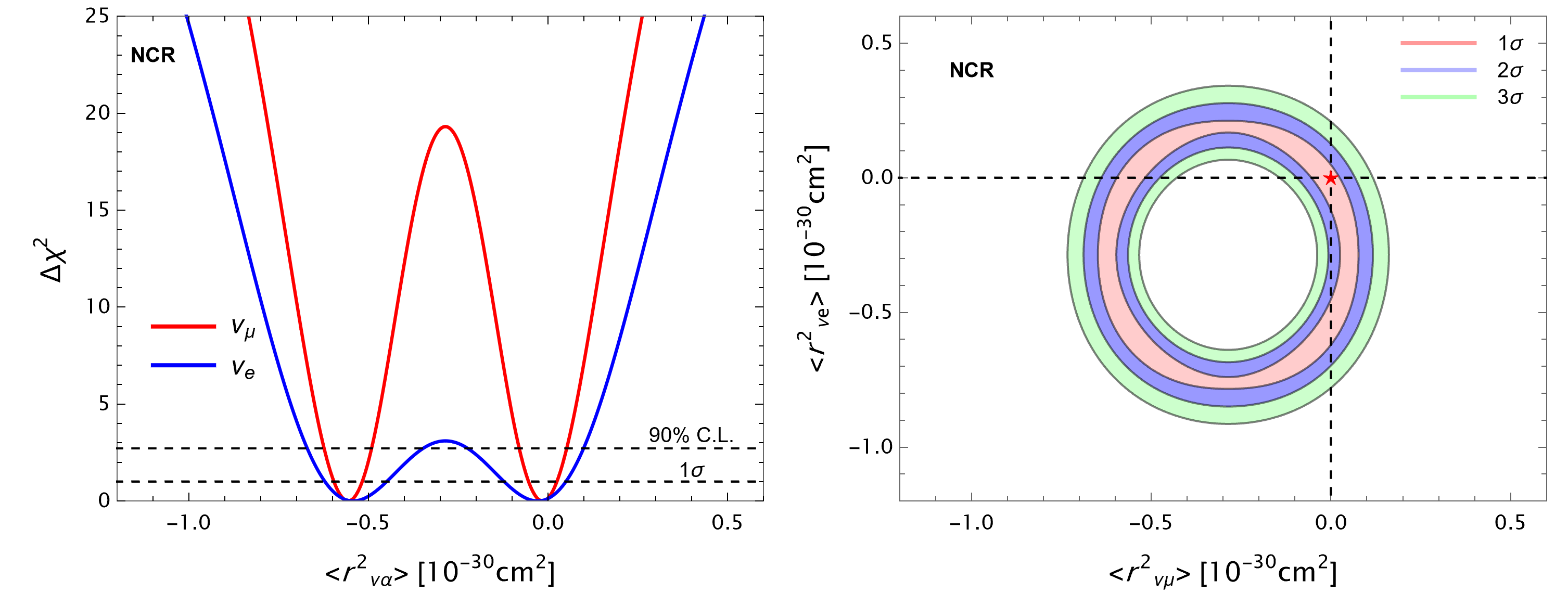}
\end{center}
\caption{Neutrino charge radius (NCR) of the electron and muon neutrino from the new COHERENT data for the one (left) and two (right) parameters fits. The red star in the right-hand side plot corresponds to the best-fit value. See text for discussion.}
\label{fig:NCR}
\end{figure}

Radiative corrections induce the neutrino charge radius for neutrinos in the SM. In the general effective electromagnetic vertex of massive neutrinos, $\bar \nu \Lambda_\mu \nu A^\mu$, the neutrino charge radius term is given by \cite{Bernabeu:2000hf,Bernabeu:2002nw,Bernabeu:2002pd},
\begin{equation}
\Lambda _{\mu }(q)=\gamma _{\mu }F(q^{2}) \simeq \gamma 
_{\mu }q^{2}\frac{\langle r^{2}\rangle }{6}\,,
\end{equation}
where $q$ is the momentum transfer and $F(q^{2}) $ is a form 
factor related to the neutrino charge radius $\langle r_{\nu}^{2}\rangle $ via 
\begin{equation}
\langle r_{\nu}^{2}\rangle =6\left. \frac{dF_{\nu }(q^{2})}{dq^{2}} \right|_{q^{2}=0} \,.
\label{eq:ncrs}
\end{equation}
The SM prediction for the charge radius of neutrino \cite{Bernabeu:2000hf,Bernabeu:2002nw,Bernabeu:2002pd,Novales-Sanchez:2013rav,Cadeddu:2018dux} is 
\begin{equation}
\langle r_{\nu_\alpha }^{2}\rangle _{\rm SM}=-\frac{G_{F}}{2\sqrt{2}\pi^2 }\left[
3-2\ln \left(\frac{m_{\alpha}^{2}}{m_{W}^{2}}\right) \right] ,
\end{equation}
where $m_{\alpha}$ is the mass of the charged lepton associated with $\nu_{\alpha}$ and $m_W$ is the mass of the $W^{\pm}$ boson. The numerical values for the corresponding flavor of neutrinos are \cite{Bernabeu:2000hf,Bernabeu:2002nw,Bernabeu:2002pd,Fujikawa:2003ww,Novales-Sanchez:2013rav,Cadeddu:2018dux} 
\begin{eqnarray}
\langle r_{\nu_e}^{2}\rangle _{\rm SM}=-0.83 \times 10^{-32} \ \ \rm cm^{2} , \nonumber \\
\langle r_{\nu_\mu }^{2}\rangle _{\rm SM}=-0.48 \times 10^{-32} \ \ \rm cm^{2} , \nonumber \\
\langle r_{\nu_\tau }^{2}\rangle _{\rm SM}=-0.30 \times 10^{-32} \ \ \rm cm^{2}.
\end{eqnarray}
In the SM, only flavor diagonal charge radii exist, while in general, transition charge radii are also possible \cite{Cadeddu:2018dux}. We consider only the former case whose contribution coherently adds to the SM cross-section. This contribution adds to the coherent cross-section by making the following replacement for the effective weak mixing angle in eq. (\ref{eq:gv&ga}), 
\begin{eqnarray}
\sin ^{2} \theta _{W}\rightarrow \sin ^{2}\theta _{W} \left(1 + \frac{\pi \alpha_{em}}{3 \sqrt{2} \sin ^{2}\theta _{W} G_F} \langle r_{\nu_\alpha }^{2}\rangle \right)
\label{eq:NCR}
\end{eqnarray}
Notice that, unlike the millicharge neutrinos in eq. (\ref{eq:milicharge}), the contribution to the cross-section due to the neutrino charge radius does not directly depend on the nuclear recoil energy and the target mass. Thus one cannot expect enhanced sensitivity at low energy recoils. This was also noted before in 
refs \cite{Khan:2020vaf, Khan:2017djo}. We fit $\langle r_{\nu_e }^{2}\rangle$ and $\langle r_{\nu_\mu }^{2}\rangle$ with the new COHERENT data first by taking one parameter at a time and then the two parameters together. We show our results in Fig. \ref{fig:NCR} and constraints from one parameter-at-a-time at 90\% C.L. are in the following.

\begin{empheq}[box=\widefbox]{align}
-0.60\times 10^{-30}<\langle r_{\nu_\mu}^{2} \rangle/{\rm cm^2} <0.05\times 10^{-30} \,,\nonumber \\
-0.67\times 10^{-30}<\langle r_{\nu_e }^{2}\rangle/{\rm cm^2} <0.10\times 10^{-30}  \,.  
\end{empheq}

\subsection{Neutrino anapole moment}
If neutrino carries a non-zero charge radius, it can also have a non-zero anapole moment \cite{zel1958electromagnetic, zel1960effect, Barroso:1984re, Abak:1987nh, Musolf:1990sa,Dubovik:1996gx, Rosado:1999yn, Novales-Sanchez:2013rav}. It determines the correlation between the spin and charge distributions of neutrinos and has the same dimensions as the charge radius. In the general vertex for electromagnetic interactions, $\bar \nu \Lambda_\mu \nu A^\mu$, the anapole term is defined by \cite{Barroso:1984re, Abak:1987nh,Rosado:1999yn}
\begin{equation}
\Lambda _{\mu }(q)=-\gamma _{\mu }\gamma _{5}F(q^{2}) \simeq -\gamma 
_{\mu}\gamma _{5}q^{2}\textit {\textbf{a}}_{\nu}\,,
\end{equation}
where the form factor ‘$F(q^{2})$’ is related to the neutrino anapole moment ‘$\textit {\textbf{a}}_{\nu \alpha}$’ by the expression,
\begin{equation}
\textit {\textbf{a}}_{\nu} =-\left. \frac{dF_{\nu }(q^{2})}{dq^{2}} \right|_{q^{2}=0} \,.
\end{equation}
By comparing with eq. (\ref{eq:ncrs}), one can write the SM prediction for the neutrino anapole moment in terms of the charge radius as \cite{Barroso:1984re, Abak:1987nh, Rosado:1999yn, Novales-Sanchez:2013rav, Cadeddu:2018dux}
\begin{equation}
\textit {\textbf{a}}_{\nu_{\rm SM}}=-\frac{\langle r_{\nu }^{2}\rangle _{\rm SM}}{6},
\end{equation} 
and numerical values accordingly are,
\begin{eqnarray}
\textit {\textbf{a}}_{\nu_{e \rm SM}}=4.98 \times 10^{-32} \ \ \rm cm^{2} , \nonumber \\
\textit {\textbf{a}}_{\nu_{\mu \rm SM}}=2.88 \times 10^{-32} \ \ \rm cm^{2} , \nonumber \\
\textit {\textbf{a}}_{\nu_{\tau \rm SM}}=1.80 \times 10^{-32} \ \ \rm cm^{2}.
\end{eqnarray}
In the case of the CE$\nu$NS, one can add the contribution of the neutrino anapole moment by replacing the effective weak mixing angle in eq. (\ref{eq:gv&ga}) as, 
\begin{eqnarray}
\sin ^{2} \theta _{W}\rightarrow \sin ^{2}\theta _{W} \left(1 - \frac{\pi \alpha_{em}}{18 \sqrt{2} \sin ^{2}\theta _{W} G_F} \textit {\textbf{a}}_{\nu_\alpha} \right)
\label{eq:NCR}
\end{eqnarray}

\begin{figure}[t]
\begin{center}
\includegraphics[width=6.5in]{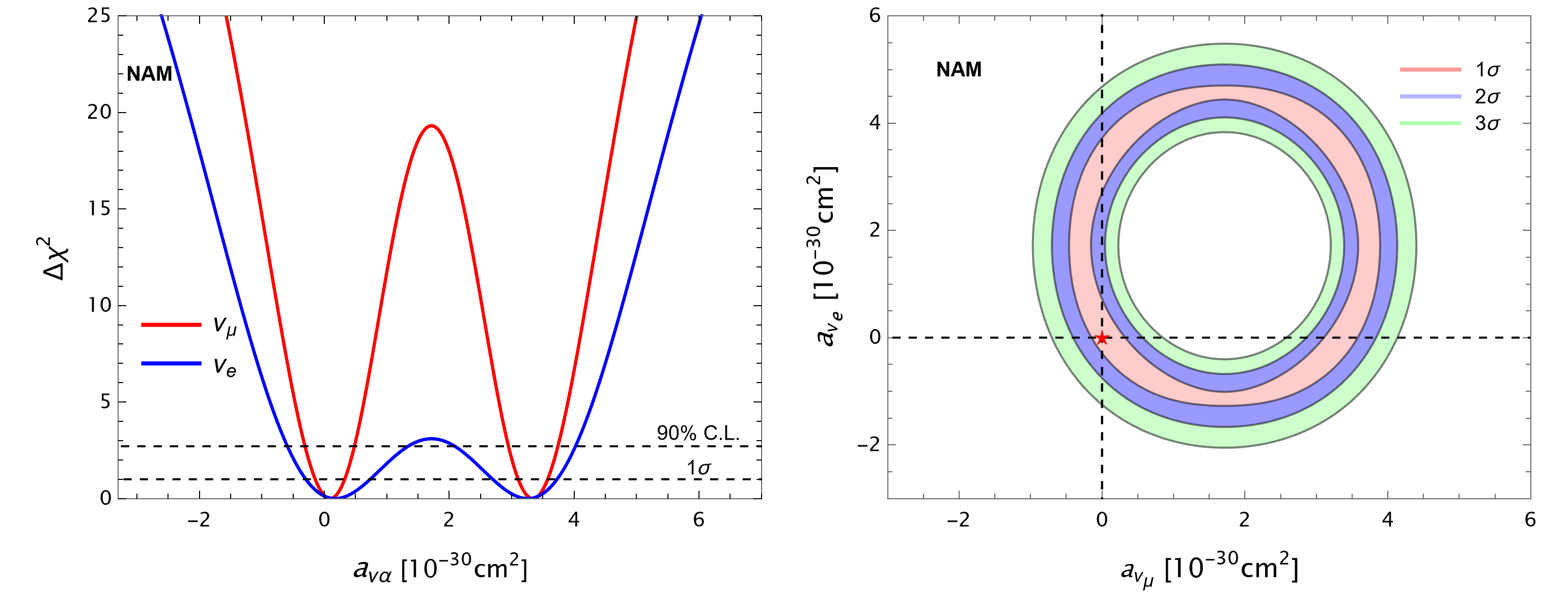}
\end{center}
\caption{Neutrino anapole moment (NAM) of the electron and muon neutrino from the new COHERENT data for the one (left) and two (right) parameters fits. The red star in the right-hand side plot corresponds to the best-fit value. See text for discussion.}
\label{fig:AM}
\end{figure}

Again one can notice that, unlike the neutrino millicharges, the anapole moment does not directly depend on the nuclear recoil energy and the target mass, and one does not expect an enhanced sensitivity at low energy recoils. We fit the parameters $\textit {\textbf{a}}_{\nu_{e \rm}}$ and $\textit {\textbf{a}}_{\nu_{\mu}}$ with the new COHERENT data. First, we fit one parameter at a time while fixing the other to zero and then fit the two parameters together. We show our obtained results in Fig. \ref{fig:AM}, and constraints from this analysis in the case of parameter-at-a-time at 90\% C.L. are the following.

\begin{empheq}[box=\widefbox]{align}
-0.30\times 10^{-30}<\textit {\textbf{a}}_{\nu_{\mu}}/{\rm cm^2} <3.7\times 10^{-30} \,,\nonumber \\
-0.60\times 10^{-30}<\textit {\textbf{a}}_{\nu_{e \rm}}/{\rm cm^2} <4.0\times 10^{-30}  \,.  
\end{empheq}

\section{Theoretical and Experimental significance of millicharged neutrinos}\label{sec:Kineconside}
The electric charge quantization comes from its empirical observation. The standard model (SM) does not predict the origin of electric charge quantization. However, several theories beyond the SM like those with magnetic monopoles \cite{Dirac:1931kp}, grand unified theories \cite{Georgi:1974sy,Pati:1974yy} and the extra dimension models \cite{Arkani-Hamed:2006emk} do predict the charge quantization. In addition, there are several extensions of the SM that predict new particles with fractional charges \cite{Ignatiev:1978xj,Okun:1983vw,Holdom:1986eq,Kors:2004dx,Batell:2005wa,Cheung:2007ut}. The fractionally charged particles are also promising candidates for dark matter \cite{Goldberg:1986nk, Mohapatra:1990vq,Kors:2005uz,Gies:2006ca,Cheung:2007ut,Feldman:2007wj,Berlin:2019uco, Berlin:2021kcm,Agrawal:2021dbo,Aboubrahim:2021ohe}. Even charges of the SM particles can deviate from the integer multiple of ‘e/3’, where ‘$e$’ is the magnitude of the unit electric charge. Neutrinos are the favorite candidates for such particles, often called milli-charged neutrinos \cite{Babu:1989tq,Babu:1989ex,Foot:1990uf,Foot:1992ui}.

The SM is an anomaly-free gauge theory with several accidental global symmetries. In the SM with one generation of fermions, neutrinos are neutral because all the SM anomalies consistently cancel out. However, for the SM with three generations, at least two of the three massless neutrinos could be electrically charged. This dequantization leads to the emergence of three gaugeable $U(1)$ symmetries, $L_e-L_{\mu}$, $L_{\mu}-L_e$ and $L_e-L_{\tau}$. Only one of the three differences can be anomaly-free and the corresponding difference in each case adds to the hypercharge of the SM. Thus, it leads to the fractional charges of neutrinos and to the dequantization of electric charge at least in the lepton sector \cite{Foot:1990uf, Foot:1992ui}.

For massive neutrinos, the existence of fractional charges of neutrinos depends on the nature of their masses. If neutrinos are Majorana fermions, then new anomalies arise, but they cancel out, leaving neutrinos neutral \cite{Babu:1989tq, Babu:1989ex}. Consequently, the electric charge remains quantized in the minimally extended SM. On the other hand, Dirac neutrinos with three right-handed partners, singlets under $SU(3)_c \times SU(2)_L$, are electrically charged. These small electric charges of Dirac neutrinos modify the charges of charged leptons and quarks due to their hypercharges' dependence on the right-handed neutrinos' hypercharges. It leads to the dequantization of electric charges in general \cite{Foot:1990uf, Foot:1992ui}. This dequantization is related to the emergence of anomaly-free gaugeable $B-L$ symmetry \cite{Babu:1989tq}. From the theoretical considerations, there is no upper limit available on the millicharge neutrinos. All the known limits are experimental \cite{Davidson:1991si, Babu:1993yh,Bressi:2011yfa, Gninenko:2006fi,Chen:2014dsa,TEXONO:2018nir,Khan:2019cvi,Cadeddu:2019eta,Cadeddu:2020lky,Khan:2020vaf} or observational \cite{Barbiellini:1987zz, Raffelt:1999gv, Davidson:2000hf,Melchiorri:2007sq,Studenikin:2012vi}. 

Neutrinos are electrically neutral in the SM. In several extensions of SM with Dirac neutrinos, the electric charges of the neutrino can arbitrarily take any values. Currently, all information about it comes from experiments. No theoretically predicted value is available. More importantly, the limits strongly depend on the process under consideration. Three critical factors contribute to neutrino millicharges. These are the interference of their amplitude with the standard model, inverse power dependence on the recoil energy, and the size of the target mass.
On the other hand, the other electromagnetic properties partially depend on these factors. For instance, there is the interference of the neutrino charge radius and anapole moment with the SM, but there is no inverse recoil energy and target mass dependence. Likewise, for the neutrino magnetic moment, there is no interference with SM and no dependence on the target mass, while there is only one power of inverse recoil energy dependence.

\begin{figure}[t]
\begin{center}
\includegraphics[width=6.5in]{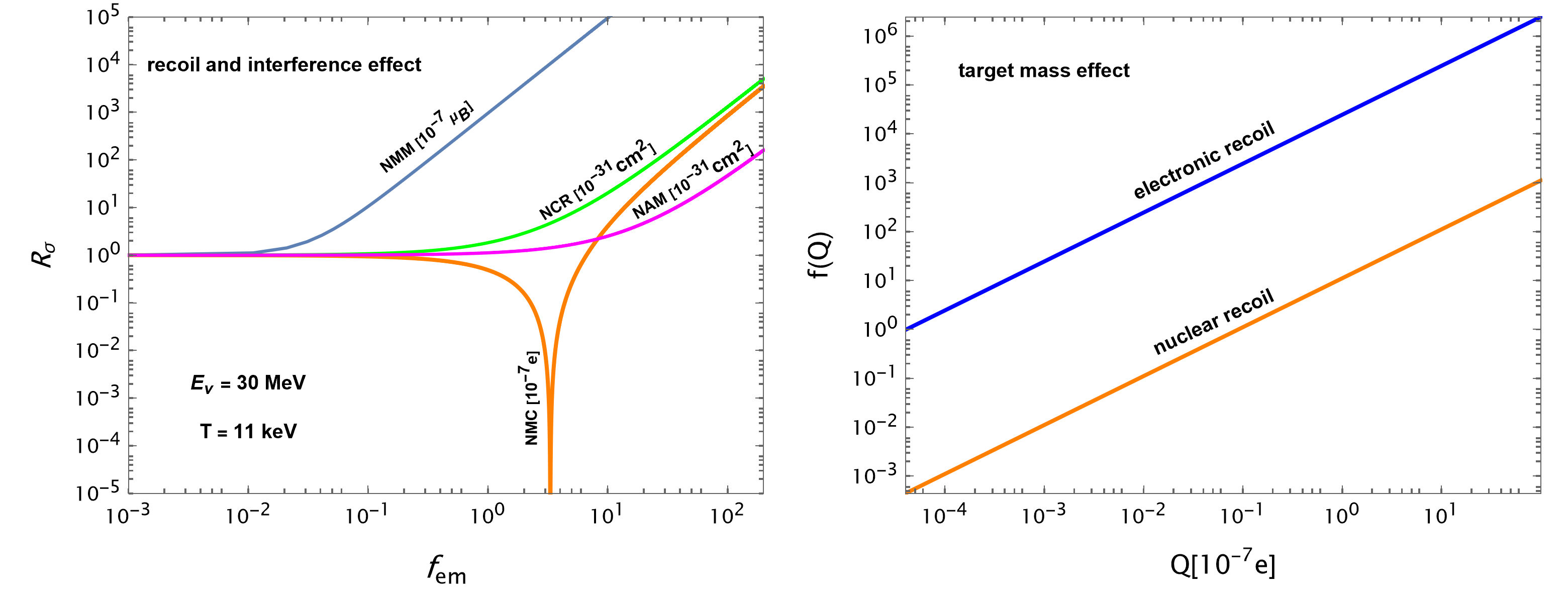}
\end{center}
\caption{Effects from interference and kinematics of the process. $\bold{Left:}$ The ratio between the millicharge, magnetic moment, charge radius, anapole moment cross-sections plus the SM cross-sections and the SM cross-section as a function of the millicharge, magnetic moment, charge radius, and anapole moment at fixed neutrino energy and nuclear recoil energy. $\bold{Right:}$ The millicharge term in Eq. (\ref{eq:milicharge}) as a function of $Q$ for the target nuclear and electronic masses. See the text for the discussion.}
\label{fig:kine}
\end{figure}

We show the millicharge dependence on the interference and the inverse power of the recoil energy for the relevant fixed neutrino energy (30 MeV) and recoil energy (11 keV) in the left-hand side plots of Fig. \ref{fig:kine}. Here, the ratio between the new physics plus the SM differential cross-section and the SM alone cross-section for the four cases was taken as a function of the neutrino millicharge, magnetic moment, charge radius, and anapole moment. In the region of interest for CE$\nu$NS, millicharge interactions compete with the SM cross-section up to $10^{-7} e$ and drop to zero $3.3 \times 10^{-7} e$ at the start and start growing afterward. In contrast, the magnetic moment starts deviation from the SM at $10^{-9} \mu_B$, the charge radius starts deviation from the SM at $10^{-31} cm^2$ and anapole moment starts deviation from the SM at around $10^{-31} \rm cm^2$. 

We plot the second term of eq. (\ref{eq:milicharge}), writing as 
\begin{eqnarray}
f\left(Q\right)=\left(\frac{\pi \alpha_{em}}{\sqrt{2} \sin ^{2}\theta _{W} G_F MT}Q_{\alpha\alpha}\right),
\label{eq:milicharge1}
\end{eqnarray}
to show the target mass dependence as a function of millicharge with target mass dependence. The comparison between a CsI nuclear target and any electronic target is shown in the right-hand side plot of Fig. \ref{fig:kine}. The millicharge contribution is suppressed at  $\mathcal{O} (10^{3})$ due to the nuclear mass in comparison to electric recoils. This effect gets relatively weaker at the cross-section level because the SM CE$\nu$NS cross-section is directly proportional to the target mass times the sum of the squares of the total number of protons and neutrons. However, the electronic targets still dominate, and there are at least two orders of magnitude intrinsically stronger constraints on millicharge neutrinos in this case than the CE$\nu$NS. Notice that this property does not hold for the other three electromagnetic properties because of no inverse dependence on the target mass. Thus, the difference in the limits strongly depends on the difference between the electronic and the nuclear masses, no matter how different the precision of the two types of experiments is.
\begin{table*}[t]
\begin{center}
\begin{tabular}{c|c|c|c|c}
\hline \hline
Flavor & $|\mu _{\nu }|[\times 10^{-11}\mu _{B}]$ &  $Q_{v}\ [\times 10^{-13}e]$& $\left \langle r_{\nu}^{2}\right \rangle \ [\times 10^{-32}$cm$^{2}]$
& $a_{\nu }\ [\times 10^{-32}$cm$^{2}]$
\\ \hline 
$\nu_e\ ($COH-2021$)$ & $ \leq 400 $ & $\  \ [-1.10,\ 3.90] \times 10^{6}\ $& $\  [-67, 10] $  & $[-60, 400] $ \\
$\nu _{\mu }($COH-2021$)$ & $ \leq 40 $ & $[-0.55,\ 0.75] \times 10^{6}
\ $&  $ [-60, 5] $ & $[-30, 370] $ \\ 
$\nu _{\tau }($COH-2021$)$ & $ - $ & $-$ & $-$   & $-$ \\ \hline \hline
$\nu_e\ ($COH-Prev$)$ & $ \leq 860 $ & $\  \ [-2.36,\ 8.38] \times 10^{6}\ $& $\  [-55, 13] $  & $[-78, 330] $ \\
$\nu _{\mu }($COH-Prev$)$ & $ \leq 570 $ & $[-1.18,\ 1.60] \times 10^{6}
\ $&  $ [-50, 8] $ & $[-48, 300]$ \\ 
$\nu _{\tau }($COH-Prev$)$ & $ - $ & $-$ & $-$   & $-$ \\ \hline \hline
$\nu _{e}\ ($Others$)$ & \multicolumn{1}{|l|}{$%
\begin{array}{l}
\leq 0.63\ \text{(XENONnT)} \ \\
\leq 3.9\  \text{(Borexino)}\  \\ 
\leq 110\  \text{(LAMPF)}\  \\ 
\leq 11\   \text{(Super-K)}\ \\ 
\leq 7.4\  \text{(TEXONO)} \\ 
\leq 2.9\  \text{(GEMMA)}\
\end{array}%
$} & $%
\begin{array}{l}
[-1.3,\ 4.7] \ \text{(XENONnT)} \\
\leq 15 \ \ \text{(Reactor)} \\ \\ \\ \\ \\
\end{array}%
\ $  & $\  \ 
\begin{array}{l}
[-45,\ 3.0] \ \text{(XENONnT)} \\
\lbrack 0.82,\ 1.27]\  \text{(Solar)} \\ 
\lbrack -5.94,\ 8.28]\  \text{(LSND)} \\ 
\lbrack -4.2,\ 6.6]\  \  \text{(TEXONO)} \\ \\ \\
\end{array}%
$ & $\begin{array}{l}
[-23,65] \\ \text{(XENONnT)} \\ \\
\end{array}$ \\ \hline
$\nu _{\mu }($Others$)$ &\multicolumn{1}{|l|}{$%
\begin{array}{l}
\leq 1.37\ \text{(XENONnT)}\ \\
\leq 5.8\  \text{(Borexino)}\ \\ 
\leq 68\  \text{(LSND)}\ \\ 
\leq 74\  \text{(LAMPF)}\
\end{array}%
$} & $[- 8.9,\ 8.6]$ \text{(XENONnT)} & \multicolumn{1}{|l|}{$%
\begin{array}{l}
[-45,\ 52] \ \text{(XENONnT)} \  \\ 
\lbrack -9,\ 31]\  \text{(Solar)} \\ 
\leq 1.2\  \text{(CHARM-II)} \\ 
\lbrack -4.2,\ 0.48]\  \text{(TEXONO)}%
\end{array}%
$}  & $\begin{array}{l}
[-95, 89] \\ \text{(XENONnT)} \\ \\ 
\end{array}$ \\ \hline
$\nu _{\tau }($Others$)$ & \multicolumn{1}{|l|}{$%
\begin{array}{l}
\leq 1.24\ \text{(XENONnT)} \ \\
\leq 5.8\  \text{(Borexino)} \\ 
\leq 3.9\times 10^{4}\ \text{(DONUT)}%
\end{array}%
$} &  $%
\begin{array}{l}
[-7.9,\ 7.8] \ \text{(XENONnT)} \  \\ 
\leq 10^{-8}\ \text{(Beam dump)} \\ \\
\end{array}%
$  &$
\begin{array}{l}
[-40,\ 45] \ \text{(XENONnT)} \\
\lbrack -9,\ 31]\  \text{(Solar)} \\ \\
\end{array}%
$  & $\begin{array}{l}
[-86, 79] \\ \text{(XENONnT)} \\ \\
\end{array}$ \\ \hline \hline
\end{tabular}%
\\[0pt]
\end{center}
\caption{90\% C.L. bounds on neutrino magnetic moment, neutrino millicharges, charge radius and anapole moment from COHERENT new data (COH-2021), COHERENT previous (COH-prev) data \cite{Khan:2020csx}  and other laboratory experiments. Bounds for XENONOnT were taken from ref. \cite{Khan:2022bel}, Borexino from ref. \cite{Borexino:2017fbd}, and solar from ref. \cite{Khan:2017djo} while all other bounds for other experiments were taken from ref. \cite{Giunti:2015gga}.}
\label{tabel1}
\end{table*}

\section{\label{sec:concl}Summary and Conclusions}
In this work, we have analyzed the new data from the COHERENT experiment to derive limits on four types of neutrino electromagnetic interactions: the neutrino millicharge, magnetic moment, charge radius, and neutrino anapole moment. In the latest update of COHERENT experimental on the detection of coherent neutrino-nucleus elastic scattering, the collaboration has reported almost double statistics and improved the precision to twice their first result. Furthermore, the statistical significance of the observed process has now enhanced to 11.6$\sigma$. This improvement motivates the improved constraints on any new physics. Finally, following our previous work, we have analyzed the new data to constrain neutrino electromagnetic interactions.

We have derived one parameter at-a-time and two-parameter bounds for the two types of neutrino flavors involved in the COHERENT experimental detection. In general, there is an improvement in constraints by a factor of two than previous constraints obtained from the COHERENT old data \cite{Khan:2019cvi}. Also, the constraints on the electromagnetic interactions of muon neutrino flavor are stronger than those on the electron flavor because of the larger fluxes and, thus, larger statistics of the muon neutrino flavors. 

For comparison, we summarize limits from the COHERENT previous data, \cite{Akimov:2017ade} and neutrino-electron elastic scattering from several other experiments including reactor, solar, and accelerator in Table \ref{tabel1}. One can see that as expected the bounds from the COHERENT new data have improved by a factor of two to the previous bounds. On the other hand, as discussed before, the limits from neutrino-electron scattering are stronger by several orders of magnitude in most cases. The limits from the recent XENONnT experiment are currently the strongest among all \cite{Khan:2022bel}.

We have discussed the importance of millicharge neutrinos in detail. Millicharge neutrinos are important from theoretical and observational points of view. Their observation would be strong evidence for physics beyond the SM and have significant astrophysical implications. It will support neutrinos' Dirac nature and the electric charges' dequantization in the standard model. As shown in Fig. \ref{fig:milchage}, there is a slight preference for the non-zero best value of the neutrino millicharges. This effect is more prominent in the two-parameter case of Fig. \ref{fig:milchage}. This sensitivity can be understood by the unique kinematic behavior of the millicharge neutrino interactions, as discussed in detail. \ref{sec:Kineconside}.
While the interference with the SM and unique dependence of the neutrino millicharge interaction is obvious when we have shown its dependence on the target mass by comparing the electronic recoil in fig. \ref{fig:kine}. We showed that electronic targets could provide intrinsically stronger limits than nuclear recoils, while their sizes depend on the specific interaction process. 

In conclusion, we have obtained new constraints on all electromagnetic interactions of neutrinos using the latest data from the COHERENT experiment. The current and future experiments of the coherent elastic neutrino-nucleus scattering experiments for both spallation neutron or reactor neutrino sources would improve the sensitivity to electromagnetic interaction. In particular, the neutrino millicharge and neutrino magnetic moment are more significant at such lower recoils. Furthermore, a better understanding of neutrino electromagnetic interactions will help understand other nonstandard neutrino interactions and dark matter in direct detection experiments.

\textbf{\uline{Note Added:}} When this work was under review, ref. \cite{AtzoriCorona:2022qrf} appeared on arXiv which also discusses and constrains the neutrino electromagnetic properties using the same new data set of COHERENT.

\begin{acknowledgments}
\noindent 
The author thanks Kate Scholberg and Dan Pershey for providing helpful information about the data and other details. I also thank Douglas McKay (KU) for his valuable suggestions on the manuscript.  The Alexander von Humboldt Foundation financially supported this work.
\end{acknowledgments}

\bibliographystyle{apsrev4-1}
\bibliography{refs}

\end{document}